# Giant *c*-axis nonlinear anomalous Hall effect in T$_d$-MoTe$_2$ and WTe$_2$

Archana Tiwari[1], Fangchu Chen[1], Shazhou Zhong[1], Elizabeth Drueke[2], Jahyun Koo[3], Austin Kaczmarek[2], Cong Xiao[4], Jingjing Gao[5], Xuan Luo[5], Qian Niu[4], Yuping Sun[5,6,7], Binghai Yan[3], Liuyan Zhao[2], and Adam W. Tsen[1*]

[1]Institute for Quantum Computing, Department of Physics and Astronomy, and Department of Chemistry, University of Waterloo, Waterloo, Ontario N2L 3G1, Canada

[2]Department of Physics, University of Michigan, Ann Arbor, Michigan 48109, USA

[3]Department of Condensed Matter Physics, Weizmann Institute of Science, Rehovot 7610001, Israel

[4]Department of Physics, The University of Texas at Austin, Austin, Texas 78712, USA

[5]Key Laboratory of Materials Physics, Institute of Solid State Physics, Chinese Academy of Sciences, Hefei, 230031, China

[6]Anhui Province Key Laboratory of Condensed Matter Physics at Extreme Conditions, High Magnetic Field Laboratory, Chinese Academy of Sciences, Hefei, 230031, China

[7]Collaborative Innovation Center of Advanced Microstructures, Nanjing University, Nanjing, 210093, China

*Correspondence to: awtsen@uwaterloo.ca

**Abstract:**
While the anomalous Hall effect can manifest even without an external magnetic field, time reversal symmetry is nonetheless still broken by the internal magnetization of the sample. Recently, it has been shown that certain materials without an inversion center allow for a nonlinear type of anomalous Hall effect whilst retaining time reversal symmetry. The effect may arise from either Berry curvature or through various asymmetric scattering mechanisms. Here, we report the observation of an extremely large *c*-axis nonlinear anomalous Hall effect in the non-centrosymmetric T$_d$ phase of MoTe$_2$ and WTe$_2$ without intrinsic magnetic order. We find that the effect is dominated by skew-scattering at higher temperatures combined with another scattering process active at low temperatures. Application of higher bias yields an extremely large Hall ratio of $E_\perp/E_\parallel = 2.47$ and corresponding anomalous Hall conductivity of order $8 \times 10^7 \text{S/m}$.

**Introduction:**
In ferromagnetic conductors, passage of charge current can generate an electric field in the transverse direction even without the application of an external magnetic field. This anomalous Hall effect (AHE) requires broken time reversal symmetry and originates both from topological aspects of the material's band structure and electron scattering coupled to the spin-orbit interaction[1]. While parsing the different contributions of the AHE has been a grand challenge in condensed matter physics for many decades, the overall strength of the AHE can be clearly quantified by the Hall ratio. Defined as the ratio between the transverse and longitudinal conductivities or fields, it ranges between ~$10^{-2}$ in common magnets and ~$10^{-1}$ in more exotic, giant AHE compounds[2,3].

In non-centrosymmetric materials retaining time reversal symmetry, it is possible to realize a nonlinear AHE (NLAHE), whereby the current induces an effective magnetization in the sample and establishes a transverse electric field that increases quadratically with applied electric field[4–12]. Experimentally, an AC voltage of frequency $\omega$ is applied across the sample and a voltage of either frequency $2\omega$ or 0 (DC) is detected in the transverse direction. Such an effect was first demonstrated in two-dimensional (2D) WTe$_2$ within the plane of the layers[13,14], as well as in several subsequent material systems[15–18]. In contrast to the ordinary AHE, the strength of the NLAHE can be quantified by the transverse field relative to the square of the applied longitudinal field. This value was measured to be $1.4 \times 10^{-9}$ m/V for 2D WTe$_2$ at low temperature[14].

Here, we report the observation of a new thickness-dependent, $c$-axis NLAHE in the same transition metal dichalcogenide family, where application of an in-plane current generates a Hall field perpendicular to the layers[6]. We focus mainly on T$_d$-MoTe$_2$, although we also verify that WTe$_2$ exhibits qualitatively similar behavior. The $c$-axis NLAHE strength can be as large as 0.43m/V. Due to the quadratic increase of the transverse field with increasing current, a Hall ratio of $E_\perp/E_\parallel = 2.47$ can be achieved at higher bias, the largest yet reported without edge states. The corresponding anomalous Hall (transverse) conductivity, ~$8 \times 10^7$ S/m, is furthermore the largest in any material. These effects arise primarily from skew-scattering at higher temperatures combined with another scattering process at lower temperatures that increases the NLAHE strength exponentially with linear longitudinal conductivity. By performing measurements across different temperatures and sample thicknesses down to the 2D limit, we furthermore uncover a general scaling behavior of the NLAHE strength with conductivity, which we parse and analyze in the two temperature regimes.

**Results:**
Nonlinear Susceptibility Tensor and Measurement Geometry
The left side of Fig. 1a shows the crystal structure of T$_d$-MoTe$_2$. The $a$ and $b$ axes lie within the plane of the layers, while the $c$-axis points out-of-plane. In addition to a mirror plane that is normal to the $a$-axis, there is a glide plane that is normal to the $b$-axis and a two-fold screw axis along the $c$-axis, putting the bulk crystal in the non-centrosymmetric space group $Pmn2_1$[19]. The corresponding point group $mm2$ has the nonlinear susceptibility tensor:
$\begin{bmatrix} 0 & 0 & 0 & 0 & d_{15} & 0 \\ 0 & 0 & 0 & d_{24} & 0 & 0 \\ d_{31} & d_{32} & d_{33} & 0 & 0 & 0 \end{bmatrix}$[20]. For the application of an in-plane electric field of frequency $\omega$: $[E_b^\omega, E_a^\omega, 0]$, a vertical Hall current at frequency $2\omega$: $j_c^{2\omega} = d_{31}(E_b^\omega)^2 + d_{32}(E_a^\omega)^2$, develops along the $c$-axis, while nonlinear currents in-plane strictly vanish: $j_b^{2\omega} = j_a^{2\omega} = 0$. In contrast, since the glide plane and screw axis symmetries are broken for the surface layers, ultrathin samples understood to be in space group $Pm$[21], with allowed nonlinear currents both in-plane and out-of-plane.

To measure this *c*-axis NLAHE, we have fabricated samples with vertical as well as in-plane contacts. In particular, our devices consist of underlying gold electrodes with MoTe$_2$ crystals of various thicknesses transferred on top. To maintain consistency, all MoTe$_2$ flakes were exfoliated from a single piece of bulk crystal grown by the flux method (see Methods). Few-layer graphene was used as a vertical top electrode with insulating hexagonal boron nitride (h-BN) blocking all but the tip of the vertical contacts to prevent mixing with in-plane currents and fields. In order to protect the sample from degradation, the entire transfer process was performed within a nitrogen-filled glovebox, while a final layer of h-BN was used to cover the whole structure. A device schematic is shown (without the top h-BN layer for clarity) on the right of Fig. 1a and an optical image of a representative device is shown in the inset.

Since current applied along the *a*- and *b*-axis of the crystal generally give rise to different nonlinear Hall currents or fields as per the different elements of the susceptibility tensor $d_{31}$ and $d_{32}$, we have controlled for the sample orientation by selecting MoTe$_2$ flakes that were rectangular in shape and aligning them with the circular electrode pattern. This allows for current injection along both the long and short directions of the flake in the same device, which in almost all cases correspond to the *a*- and *b*-axis, respectively. For either current $I \parallel a$ or $I \parallel b$, our device geometry yields measurement of the following linear and nonlinear voltages: in-plane longitudinal ($V_{xx}$), in-plane Hall ($V_{xy}$), and out-of-plane Hall ($V_{xz}$).

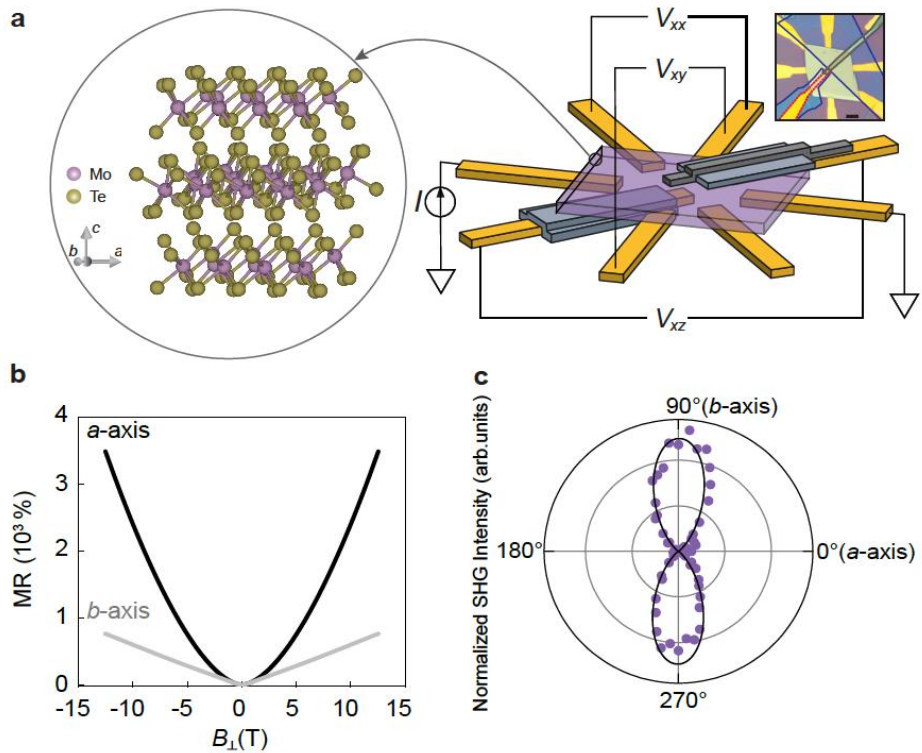

**Fig. 1. NLAHE measurement schematic and sample characterization. a,** Main panel: crystal structure of MoTe$_2$ in the T$_d$ phase and device geometry. Purple, blue, and gray flakes represent MoTe$_2$, h-BN, and graphene, respectively. Inset shows optical image of 70-nm-thick device. h-BN blocking layers, graphene, and the bottom vertical contact are outlined in blue, gray, and red, respectively. Scale bar is 5 µm. **b,** Symmetrized magnetoresistance (MR) of 47-nm-thick device with current applied along the *a* and *b* crystal axes at 1.4K. $I \parallel a$ ($I \parallel b$) exhibits quadratic (linear) MR. **c,** Angle-dependent second harmonic generation (SHG) intensity measured on the same sample. Minimum (maximum) intensity occurs along the *a*-axis (*b*-axis).

Determination of Crystalline Axes

In order to confirm the crystal orientation of the flakes, we first measured the longitudinal magnetoresistance, $\text{MR} = [V_{xx}(B_\perp) - V_{xx}(0)]/V_{xx}(0)$ for constant current, along the two current directions. Figure 1b shows representative data from a 47-nm-thick device at 1.4K: $I \parallel a$ ($I \parallel b$) exhibits quadratic (linear) magnetoresistance up to 12.5T, consistent with previous measurements on bulk crystals[22]. Overall, MR is large and non-saturating due to close electron-hole compensation combined with relatively high carrier mobilities[23–25]. In addition, we have performed optical second harmonic generation (SHG) measurements, which can also distinguish between the *a*- and *b*-axis of MoTe$_2$. Figure 1c shows SHG intensity measured on the same sample at 80K with incident and scattered light polarization in parallel as a function of the polarization angle relative to the *a*-axis, at which a node can be seen, consistent with previous results[26]. Unlike the electrical second harmonic measurements we are to present, optical SHG is mostly sensitive to the symmetry properties of the surface layers, which allow for in-plane responses (see Supplementary Note 1).

Observation of *c*-axis Nonlinear Anomalous Hall Effect

Figure 2a shows representative NLAHE measurements taken on a 127-nm-thick sample at 2K. The upper two panels show $V_{xz}$ measured at DC and second harmonic ($2\omega = 354$Hz) frequencies as a function of the first harmonic ($\omega = 177$Hz) $V_{xx}^2$ for $I \parallel a$ and $I \parallel b$. All four traces show linear behavior at low bias as expected for the NLAHE[14], although the slope is larger for current along the *a*-axis for reasons which we shall discuss. At higher bias the curves become slightly sublinear, possibly due to sample heating. Furthermore, for each current direction, the DC and $2\omega$ amplitudes are comparable for a given applied bias. In contrast, as shown in the lower two panels, relatively small second harmonic voltage is measured for $V_{xx}$ or $V_{xy}$ for comparable bias, consistent with the allowed symmetries discussed earlier for bulk-like crystals. Although, $V_{xy}^{2\omega}$ for $I \parallel a$ is in principle allowed for thin or surface layers (see Supplementary Note 2). These results do not change when the frequency $\omega$ is changed over an order of magnitude or by exchanging the current leads, indicating that they stem from intrinsic properties of the sample (see Supplementary Note 3).

For the remainder of this work, we shall focus only on the second harmonic component of the *c*-axis NLAHE. The strength of this effect can be defined as the slope of the linearized plots presented at low bias. In order to understand and demonstrate the reproducibility of these results, we have performed similar measurements across five samples of different thicknesses down to the 2D limit: 127, 70, 47, 32, and 9nm. The NLAHE strength is shown as a function of sample thickness for both $I \parallel a$ and $I \parallel b$ in the upper panel of Fig. 2b. We have used electric field values, $E_z^{2\omega}/(E_x^\omega)^2$, instead of voltage in order to account for the different dimensions of the samples. The local $E_x^\omega$ at the vertical contact area was obtained from finite element method simulations of each individual device (see Methods and Supplementary Note 4). For every sample, the *a*-axis shows larger strength than the *b*-axis, while for both axes, the strength decreases substantially with decreasing thickness. In particular, for $I \parallel a$ in the thickest (127nm) sample, $E_z^{2\omega}/(E_x^\omega)^2 = 0.43$m/V.

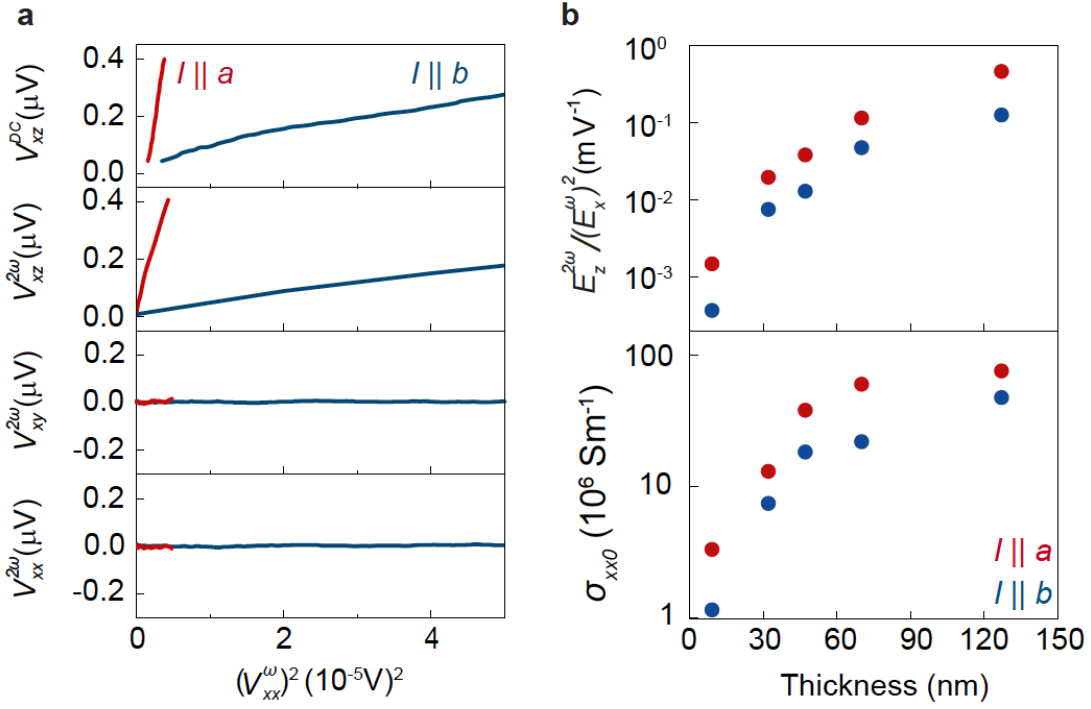

**Fig. 2. Dependence of NLAHE on crystal axis and thickness. a,** Upper panels: DC and second harmonic $V_{xz}$ vs. first harmonic $V_{xx}^2$ for $I \parallel a$ and $I \parallel b$ in 127-nm-thick sample at 2K. Lower panels: analogous second harmonic measurements for in-plane response show significantly weaker signal under comparable bias. **b,** NLAHE strength $E_z^{2\omega}/(E_x^{\omega})^2$ and residual longitudinal conductivity $\sigma_{xx0}$ vs. sample thickness for $I \parallel a$ and $I \parallel b$. Both values are larger for greater thickness and $I \parallel a$.

Scaling with Sample Conductivity
The differences observed between the samples and two crystalline axes can be attributed to their differences in conductivity. In the lower panel of Fig. 2b, we have plotted the residual longitudinal conductivity $\sigma_{xx0}$ for each sample and axis, which shows a similar trend. This reflects the three-dimensional electronic character of MoTe$_2$—decreasing thickness increases surface scattering, which lowers the residual conductivity by over an order of magnitude in our thinnest sample, consistent with previous results[25]. In Fig. 3a, we further show the temperature dependence of $E_z^{2\omega}/(E_x^{\omega})^2$ and conductivity measured for the 127-nm-thick sample—both decrease with increasing temperature. In the lower inset, we show the temperature-dependent longitudinal resistivity for $I \parallel a$ in the same sample, which conversely decreases with decreasing temperature. Specifically, the resistivity scales linearly with temperature at higher temperatures, but begins to saturate in the crossover region roughly centered at $T_L \sim 18K$.

Taken together, these results indicate that conductivity (or resistivity) may be the fundamental parameter upon which the NLAHE strength depends. To show this dependence explicitly, in the main panel of Fig. 3b, we have plotted, in log-log scale, $E_z^{2\omega}/(E_x^{\omega})^2$ for both axes of each sample versus temperature-dependent conductivity normalized to the residual conductivity, $\sigma_{xx}/\sigma_{xx0}$, and a general trend appears to emerge. Specifically, at lower conductivity (higher temperature) for a given sample and axis, $E_z^{2\omega}/(E_x^{\omega})^2$ scales closely with $\sigma_{xx}^2$ as compared with the guide-to-eye in gray. Above (below) a certain conductivity (temperature), it increases at an even faster rate. To see this crossover point more clearly, we have extracted the local slope $\upsilon$ from the data in the main panel as a function of temperature $T$ (normalized to the $T_L$ of each sample/axis, which ranges between 13-18K), yielding the relationship $E_z^{2\omega}/(E_x^\omega)^2 \sim \sigma_{xx}^{\upsilon(T)}$. The inset of Fig. 3b shows $\upsilon(T)$ for all the different traces in the same color scheme as that used in the main panel. Above $T_L$, $\upsilon$ is near 2 for every trace, while it rises continuously below this temperature. We have also performed similar measurements on bulk-like WTe$_2$ (which shares the same crystal structure) as well as MoTe$_2$ Hall bar devices for $I \parallel a$ (see Supplementary Note 5 and 6, respectively), both of which show qualitatively similar behavior. This indicates that the strength and scaling of the c-axis NLAHE is not unique to MoTe$_2$ nor a result of the circular electrode geometry.

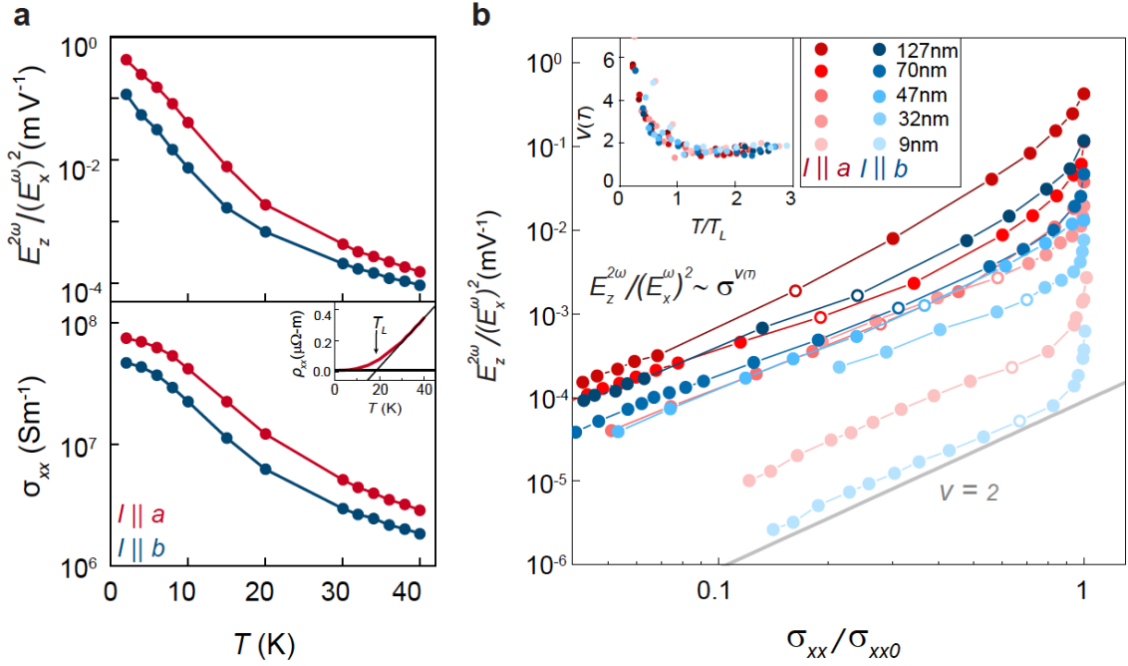

**Fig. 3. Scaling of the NLAHE with sample conductivity. a**, Main panel: $E_z^{2\omega}/(E_x^\omega)^2$ and $\sigma_{xx}$ vs. temperature for $I \parallel a$ and $I \parallel b$ in 127-nm-thick sample. Inset: longitudinal resistivity $\rho_{xx}$ vs. temperature for $I \parallel a$ in same sample. $\rho_{xx}$ crosses over from linear temperature dependence to saturation below $T_L \sim 18K$. **b**, Main panel: $E_z^{2\omega}/(E_x^\omega)^2$ vs. $\sigma_{xx}/\sigma_{xx0}$ for all four samples along $a$ and $b$ axes. All traces show scaling close to a power law with exponent of 2 at lower conductivity (higher temperature) and exhibits an upturn as $\sigma_{xx}$ approaches $\sigma_{xx0}$. Open circles mark the data points taken at the temperature closest to $T_L$. Inset: local scaling exponent $\upsilon$ vs. temperature $T$. $\upsilon$ continually increases above 2 for all traces below $T_L$.

**Discussion:**
In order to understand the mechanism behind the NLAHE and the observed scaling with conductivity, we turn to the theory behind the ordinary AHE, as it is understood that the NLAHE strength, which can also be written as $\frac{E_z^{2\omega}}{(E_x^\omega)^2} = \frac{\sigma_{AH}^{(NL)}}{\sigma_{zz}E_x^\omega}$ ($\sigma_{zz}$ being the linear c-axis conductivity and $\sigma_{AH}^{(NL)}$ is the nonlinear anomalous Hall conductivity element $zx$), scales in the same way as anomalous Hall conductivity $\sigma_{AH}$ in the ordinary (linear) AHE[9,11,14]. For the latter, $\sigma_{AH}$ consists of intrinsic Berry curvature, skew-scattering, and side-jump(-like) scattering contributions[1]. The intrinsic part does not depend on longitudinal conductivity (scales with $\sigma_{xx}^0$), while upon changing temperature, skew-scattering scales with $\sigma_{xx}^2$ and side-jump contains terms that scale with $\sigma_{xx}^0$, $\sigma_{xx}^1$, and $\sigma_{xx}^2$[9,11,27,28]. It is therefore not possible to distinguish between the various mechanisms by scaling analysis a priori. Nonetheless, it is generally understood that skew-scattering dominates for highly conducting samples with large residual conductivity[1]. Our MoTe$_2$ samples show a conductivity comparable to the most highly conducting AHE systems, and so we shall assume that the $v = 2$ scaling at higher temperatures is attributed primarily to skew-scattering.

To further our understanding, we have fit all our data for $E_z^{2\omega}/(E_x^\omega)^2$ above $T_L$ to the functional form: $A\sigma_{xx}^2 + B$, and the extracted $A$ and $B$ values are plotted in the lower panels of Fig. 4a as a function of the residual conductivity $\sigma_{xx0}$ of the sample/axis. The upper panel shows a representative fitting for the 127-nm-thick sample with $I \parallel a$. The constant term $B$ is expected to contain contributions both from the intrinsic Berry curvature and extrinsic side-jump(-like) events and be independent of residual conductivity[11,27,28]. Although this value varies over two orders of magnitude between samples, we indeed do not observe any clear trend with $\sigma_{xx0}$. We have further calculated the upper limit for the predicted intrinsic contribution to $B$ based on previous theory (see Supplementary Note 7)[6], and the result is marked by the gray line. This theoretical value is one order of magnitude less than our smallest experimental value, suggesting that intrinsic Berry curvature only plays a minor role in the large NLAHE observed here. Although, the intrinsic contribution may in principle be enhanced by external perturbations, such as strain, or symmetry-breaking by the surface layers[29–31].

On the other hand, $A$ initially decreases with increasing $\sigma_{xx0}$, but saturates at higher $\sigma_{xx0}$. For the AHE, the skew-scattering term is given by $A = \alpha\sigma_{xx0}^{-1}$[9,11,27,28], where $\alpha$ is termed the skew-scattering coefficient and the residual conductivity is taken as a measure of the disorder from impurities. This general scaling cannot be directly applied for our samples, since we are not tuning the impurity concentration by changing thickness, while measurements along the different crystalline axes for the same sample show different residual conductivity due to differences in band structure, not disorder. By reducing thickness, we have instead increased the contribution of surface scattering to the total resistivity, which by Matthiessen's rule can be written as: $\sigma_{xx0}^{-1} = \sigma_{xx0b(ulk)}^{-1} + \sigma_{xx0s(urface)}^{-1}$. Taking into account the different symmetries for the bulk and surface layers, we further assign different skew scattering coefficients between the two and write: $A = \alpha_b\sigma_{xx0b}^{-1} + \alpha_s\sigma_{xx0s}^{-1} = \alpha_s\sigma_{xx0}^{-1} + (\alpha_b - \alpha_s)\sigma_{xx0b}^{-1}$. In the thick limit, we expect $\sigma_{xx0} \sim \sigma_{xx0b}$, and so $A \sim \alpha_b\sigma_{xx0b}^{-1}$, consistent with the nonzero constant value we observe at high conductivity. If we further approximate that $\sigma_{xx0b}$ is independent of thickness and equal to the measured $\sigma_{xx0}$ of our thickest sample, we can fit our data across the entire thickness (residual conductivity) range for each axis to the formula above (see colored lines in the middle panel of Fig. 4a). The extracted skew-scattering coefficients are shown in the inset. Overall, $\alpha$ is larger for bulk skew-scattering and similar between the two axes.

For temperatures below $T_L$, $E_z^{2\omega}/(E_x^\omega)^2$ increases further with conductivity beyond the $\sigma_{xx}^2$ dependence seen at higher temperatures (see Fig. 3b). Such behavior has not been previously observed experimentally for the AHE, and so it remains an open theoretical question as to the microscopic scattering mechanism responsible for this effect. A possible candidate is side-jump, as it has been recently predicted that, for high-purity samples below the Debye or Bloch-Gruneisen temperature, scaling parameters for side-jump(-like) contributions may actually be temperature-dependent[11]. We proceed to fit our data in this regime in order to determine an empirical scaling relationship. The continuous increase of $v$ below $T_L$ suggests an exponential dependence on $\sigma_{xx}$, and so we have fit the additional contribution to: $\Delta \equiv \frac{E_z^{2\omega}}{(E_x^\omega)^2} - (A\sigma_{xx}^2 + B) = C\exp(D\sigma_{xx})$, and the extracted $C$ and $D$ values are plotted in the middle panel of Fig. 4b as a function of the residual conductivity $\sigma_{xx0}$ of the sample/axis. The upper panel shows a representative fitting for the 127-nm-thick sample with $I \parallel a$. While $C$ sharply with increasing $\sigma_{xx0}$, $D$ decreases. The net effect, however, is such that the NLAHE strength at low temperature is increased by a roughly constant factor above that expected from skew-scattering alone. In the lower panel of Fig. 4b, we have plotted as a function of $\sigma_{xx0}$, the ratio between the measured $E_z^{2\omega}/(E_x^\omega)^2$ value at 2K and $A\sigma_{xx0}^2 + B$, the extrapolated skew-contribution, and this factor falls between 3–5 for all but one sample/axis.

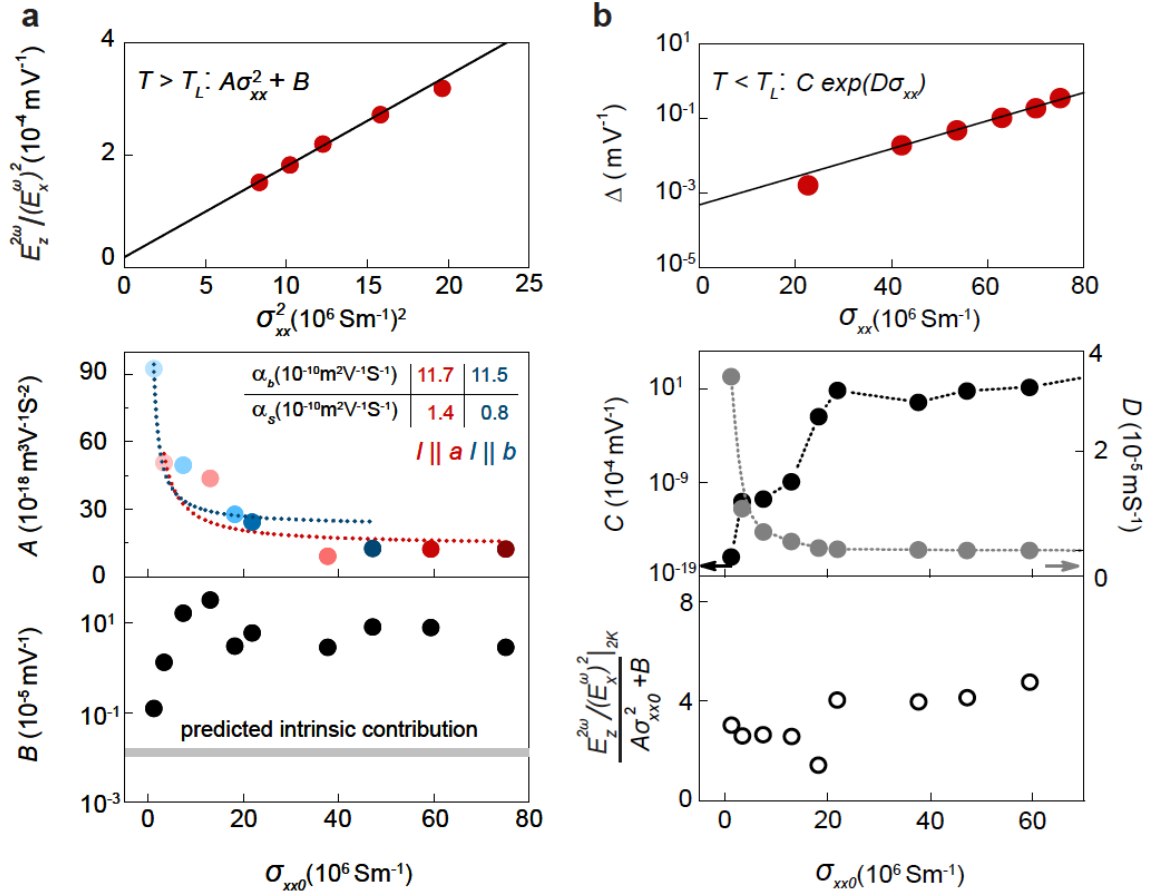

**Fig. 4. Determination and analysis of scaling parameters in NLAHE. a,** Upper panel: representative fit to $E_z^{2\omega}/(E_x^\omega)^2 = A\sigma_{xx}^2 + B$ for $T > T_L$ for 127-nm-thick sample, $a$-axis. Lower panels: $A$ and $B$ vs. $\sigma_{xx0}$ for all samples and axes. Red and blue dashed lines are fits to formulas described in the text for a and b axes, respectively. Gray line is predicted intrinsic Berry curvature contribution to $B$. **b,** Upper panel: representative fit to $\Delta = C\exp(D\sigma_{xx})$ for $T < T_L$ for 127-nm-thick sample, $a$-axis. Middle panel: $C$ and $D$ vs. $\sigma_{xx0}$ for all samples and axes. Lower panel: low-temperature enhancement factor $\frac{E_z^{2\omega}}{(E_x^\omega)^2}\Big|_{2K} / (A\sigma_{xx0}^2 + B)$ vs. $\sigma_{xx0}$.

The combination of the different scattering processes, together with the nonlinear dependence of the Hall field on applied bias, allows us to obtain the largest Hall ratio in electric field observed to date outside the quantum (anomalous) Hall regime. In Fig. 5a, we have plotted $E_z^{2\omega}/E_x^\omega$, the ratio between the Hall field generated and the applied longitudinal field, as a function of $E_x^\omega$ measured for the 127-nm-thick sample at 2K, which shows the largest NLAHE strength amongst all our samples. The ratio is always larger for $I \parallel a$, for which it rises initially with increasing $E_x^\omega$, reaching a maximum of 2.47, an order of magnitude larger than the analogous quantity measured in giant linear AHE compounds[2,3]. At larger bias, $E_z^{2\omega}/E_x^\omega$ decreases as $E_x^\omega$ itself becomes sublinear with applied current (see Supplementary Note 8), likely due to heating.

We would further like to quantify the nonlinear anomalous Hall conductivity at this peak Hall ratio, which can be expressed as: $\sigma_{AH}^{(NL)} = \frac{E_z^{2\omega}}{E_x^\omega}\sigma_{zz}$. We have measured both the *a*- and *c*-axis resistivity of a separate bulk MoTe$_2$ crystal (grown under the same conditions as that used for our devices) and further calculated the conductivity anisotropy as a function of the position of the Fermi level using density functional theory (see Supplementary Note 9). $\sigma_{zz}/\sigma_{xx} \sim 0.25 (\text{exp.}) - 0.6(\text{th.})$ for $I \parallel a$ near the charge neutrality point where electrons and holes compensate, yielding $\sigma_{AH}^{(NL)} \sim 8 \times 10^7 \text{S/m}$. Assuming a similar anisotropy ratio for WTe$_2$, we estimate $\sigma_{AH}^{(NL)} \sim 5 \times 10^7 \text{S/m}$ at the peak Hall ratio for our bulk-like WTe$_2$ device (see Supplementary Note 5). For comparison, we have plotted these values in Fig. 5b, together with those measured in-plane for ultrathin MoTe$_2$ and WTe$_2$, as well as the linear anomalous Hall conductivities measured in various magnetic systems as a function of the residual longitudinal conductivity[2,3,39–43,14,32–38]. Overall, $\sigma_{AH}$ is larger in more highly conducting systems. The values for the *c*-axis response in MoTe$_2$ and WTe$_2$, in particular, are the largest yet reported.

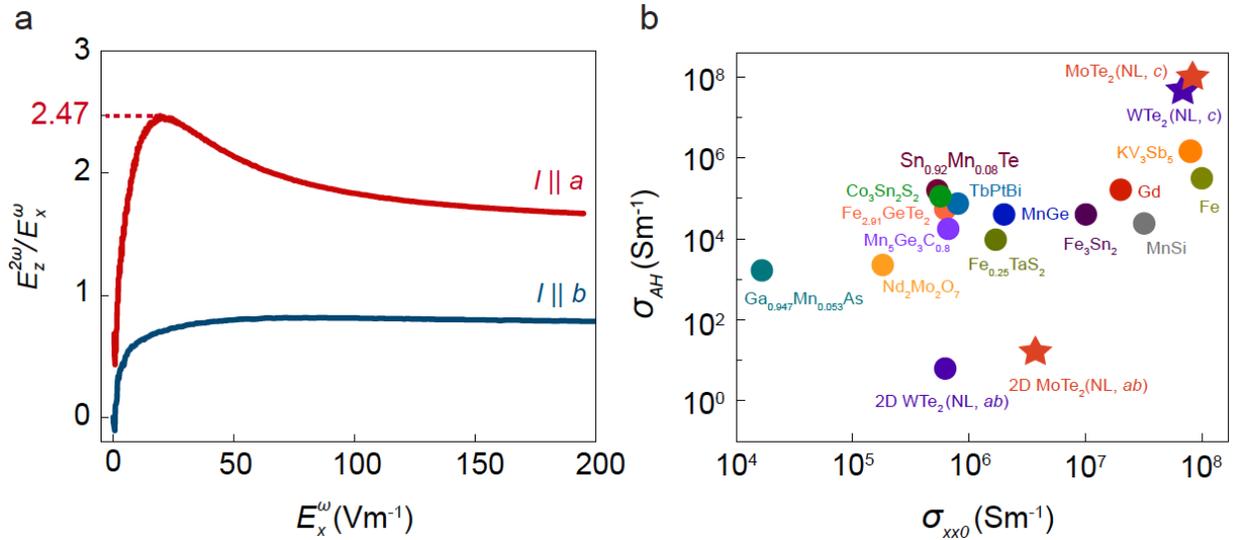

**Fig. 5. Observation of extremely large Hall ratio and conductivity at higher bias. a,** Hall ratio $E_z^{2\omega}/E_x^\omega$ vs $E_x^\omega$ for $I \parallel a$ and $I \parallel b$ in 127-nm-thick sample. Hall ratio reaches 2.47 for *a*-axis. **b,** Anomalous Hall conductivity $\sigma_{AH}$ vs. $\sigma_{xx0}$ for various linear and nonlinear (NL) AHE materials. Stars mark values measured in this study.

In conclusion, we have observed an extremely large *c*-axis NLAHE in T$_d$-MoTe$_2$ and WTe$_2$ that is dominated by asymmetric electron scattering. Our work reaffirms the importance of extrinsic contributions to the (NL)AHE, especially in highly conducting metals, and opens a new direction for obtaining giant Hall ratios and conductivities in non-centrosymmetric systems without breaking time reversal symmetry.

**Methods:**
Crystal Synthesis
1T'-MoTe$_2$ (T$_d$-WTe$_2$) single crystals were grown by the flux method using Te as a solvent. Mo (W) [Alfa Aesar, 99.9%], Te [Alfa Aesar, 99.99%] powders were ground and placed into alumina crucibles in a ratio of 1:25 (1:99) and sealed in a quartz ampoule. After the quartz ampoule was heated to 1050°C (1000°C) and held for two days (10 hours), the ampoule was slowly cooled to 900°C (460°C) over 120 hours (100 hours) and centrifuged. Shiny and plate-like crystals with lateral dimensions up to several millimeters were obtained.

Device Fabrication
Au (45nm)/Ti (5nm) electrodes in a circular geometry were pre-patterned on Si wafers with 285-nm-thick $SiO_2$ using conventional photolithography and electron-beam deposition. Graphene (Coorstek), h-BN (HQ graphene), and (Mo/W)$Te_2$ flakes were exfoliated onto blank $SiO_2$/Si wafers. After the desired flakes were identified, a polymer stamp coated with polycarbonate was used to sequentially pick up the entire h-BN/graphene/h-BN/(Mo/W)$Te_2$/h-BN heterostructure to avoid contamination between the layers. The heterostructure was then aligned and transferred onto the pre-patterned electrodes. The entire exfoliation and transfer process were performed within a nitrogen-filled glovebox to avoid degradation of (Mo/W)$Te_2$ in air.

Transport Measurements
Both the magnetotransport and NLAHE measurements were primarily carried out in a pumped Helium-4 cryostat with a base temperature of 1.4K. The latter was further cross-checked in an optical cryostat with base temperature of 5K. For the second harmonic NLAHE measurements, an AC current with frequency between 17 to 277Hz was passed along either the *a*- or *b*-axis of the (Mo/W)$Te_2$ crystal, and $V_{xx}$, $V_{xy}$, and $V_{xz}$ voltages were measured at both the first harmonic (X channel) and second harmonic (Y channel) frequencies using an SR830 lock-in amplifier. A Keithley 2450 source measure unit was further used to measure the DC voltage response of the NLAHE.

Optical Second Harmonic Generation Measurements
Rotational anisotropy SHG measurements were taken in a normal incidence geometry using a pulsed laser with a pulse duration of ~325fs, a repetition rate of 200kHz, and an incoming fundamental wavelength of 800nm. The MoTe$_2$ samples were held at 80K inside an optical cryostat. Using a single-photon sensitive detector, the intensity of the reflected SHG was measured as a function of the angle between the incident polarization and the *x*-axis in the lab coordinate frame. The incident fundamental and the reflected SHG polarizations can be selected to be either parallel or crossed, forming two polarization channels for the SHG measurements.

Finite Element Method Simulations
Simulations of the potential and current distributions in our devices were carried out using the static current conduction solver in Elmer, an open source multiphysical simulation software utilizing the finite element method. For each circular device, the geometry of the (Mo/W)$Te_2$ flake and electrodes were constructed and meshed using FreeCAD and imported into Elmer. For a given direction and magnitude of current, the MoTe$_2$ conductivity tensor was iteratively adjusted to match the potential difference measured across the longitudinal voltage leads, $V_{xx}$. The local longitudinal electric field at the vertical contacts, $E_x$, was then extracted from the simulation.

**Data Availability:**
The datasets generated during and/or analyzed during the current study are available from the corresponding author on reasonable request.

**Acknowledgements:**
We thank Kin Fai Mak for helpful discussions and a critical reading of our manuscript, Christopher Gutierrez for assistance with the device schematic, and Peng Li for assistance with initial measurements. AWT acknowledges support from the US Army Research Office (W911NF-19-10267), Ontario Early Researcher Award (ER17-13-199), and the National Science and Engineering Research Council of Canada (RGPIN-2017-03815). This research was undertaken thanks in part to funding from the Canada First Research Excellence Fund. LZ acknowledges the support by the National Science Foundation (NSF) CAREER Award (DMR-1749774). ED acknowledges support by the NSF Graduate Research Fellowship Program (DGE-1256260). BY acknowledges financial support by the Willner Family Leadership Institute for the Weizmann Institute of Science, the Benoziyo Endowment Fund for the Advancement of Science, the Ruth and Herman Albert Scholars Program for New Scientists, and the European Research Council under the European Union's Horizon 2020 research and innovation programme (Grant No. 815869). XL and YPS thank the support of the National Key Research and Development Program under contract 2016YFA0300404, the National Nature Science Foundation of China under contracts 11674326, 11874357, and the Joint Funds of the National Natural Science Foundation of China and the Chinese Academy of Sciences' Large-Scale Scientific Facility under contracts U1832141, U1932217. CX and QN were supported by the NSF (EFMA-1641101) and Welch Foundation (F-1255).

**Author Contributions:**
AT and AWT conceived and designed the experiments. FC and JG grew and characterized the $MoTe_2$ and $WTe_2$ crystals under the guidance of XL and YPS. AT fabricated the devices and performed the transport measurements. SZ performed the finite element method simulations. ED and AK performed the optical second harmonic generation measurements under the guidance of LZ. JK performed the density functional theory calculations under the guidance of BY. AT and AWT analyzed the data together with CX and QN and wrote the manuscript with input from all authors.

**Competing Interests Statement:**
The authors declare no competing interests.